\documentclass[12pt]{article}

\usepackage{amsmath}
\usepackage{epsfig}
\usepackage{amssymb}
\usepackage[colorlinks=true,
            linkcolor=blue,
            citecolor=blue]{hyperref}
\usepackage{bbold}
\usepackage{braket}
\input{epsf}
\usepackage[english]{babel}

\usepackage[numbers,sort&compress]{natbib}

\newcommand{\jet}{\text{jet}}

\def\to{\rightarrow}

\newcommand\as{\alpha_s}
\def\ep{\varepsilon}
\def\nn{\nonumber}
\def\beq{\begin{equation}}
\def\eeq{\end{equation}}
\def\beeq{\begin{eqnarray}}
\def\eeeq{\end{eqnarray}}
\def\beal{\begin{align}}
\def\eeal{\end{align}}

\begin{document}

\begin{titlepage}
\renewcommand{\thefootnote}{\fnsymbol{footnote}}
\par \vspace{10mm}

\begin{center}
\vspace*{10mm}
{\Large \bf
Access to Photon Fragmentation Functions \\[2mm] 
in Hadronic Jet Production}
\end{center}

\par \vspace{2mm}
\begin{center}
{\bf Tom Kaufmann${}^{\,a}$,}
\hskip .2cm
{\bf Asmita Mukherjee${}^{\,b}$,}
\hskip .2cm
{\bf Werner Vogelsang${}^{\,a}$  }\\[5mm]
\vspace{5mm}
${}^{a}$ Institute for Theoretical Physics, T\"ubingen University, 
Auf der Morgenstelle 14, \\ 72076 T\"ubingen, Germany\\[2mm]
${}^{b}$ Department of Physics, Indian Institute of Technology Bombay, \\
Powai, Mumbai 400076, India
\end{center}

%%%%%%%%%%%%%%%%%%%%%%%%%%%%%%%%%%%%%%%%%%%%%%%%%%%%%%%%
%%%%%%%%%%%%%%%%%%%%%%%%%%%       ABSTRACT      %%%%%%%%%%%%%%%%%%%
%%%%%%%%%%%%%%%%%%%%%%%%%%%%%%%%%%%%%%%%%%%%%%%%%%%%%%%%

\vspace{9mm}
\begin{center} {\large \bf Abstract} \end{center}
We argue that the process $pp \to (\jet \gamma)X$, for which a photon is observed inside a fully reconstructed jet 
and is treated as part of the jet, offers new probes of the so far little known fragmentation functions for photons. 
We present a next-to-leading-order QCD calculation of the cross section for this process in the limit that the
jet is relatively narrow. We also investigate the background resulting from the two-photon decay of neutral pions. 
We present numerical results relevant for possible measurements at the LHC and at RHIC. These 
suggest that $pp \to (\jet \gamma)X$ should provide clean access to the photon fragmentation functions, 
provided an efficient suppression of the background is available in experiment.

\end{titlepage}  

\setcounter{footnote}{2}
\renewcommand{\thefootnote}{\fnsymbol{footnote}}

%%%%%%%%%%%%%%%%%%%%%%%%%%%%%%%%%%%%%%%%%%%%%%%%%%%%%%%%%%%
%%%%%%%%%%%%%%%%%%%%%%%%%%%       INTRODUCTION       %%%%%%%%%%%%%%%%%%%
%%%%%%%%%%%%%%%%%%%%%%%%%%%%%%%%%%%%%%%%%%%%%%%%%%%%%%%%%%%

\section{Introduction}
The production of photons with high transverse momentum $p_T$ in hadronic collisions is
of fundamental importance in today's particle and nuclear physics. Foremost, it may serve as a
tool for determining the gluon distributions of the scattering hadrons, thanks to the presence and 
dominance of the leading order (LO) Compton subprocess $qg\to \gamma q$. Photons also
provide sensitive probes of the medium produced in collisions of heavy ions, being able to 
traverse and escape the medium with little attenuation. Finally, photon signals play an important role
in studies of physics within and beyond the Standard Model, with the process 
$pp\to \gamma \gamma X$ through production and decay of a Higgs boson
and the recent indication of a 750 GeV diphoton excess seen by the ATLAS and CMS 
collaborations \cite{diphoton} at the LHC arguably being among the most well-known examples.

When used in searches of new physics phenomena, photon ``signal'' processes 
are invariably affected by backgrounds from more mundane Standard Model sources.
In view of this it is clear that a good quantitative theoretical understanding of photon production
in high-energy hadronic collisions is crucial. 
For photons produced in a process characterized by a 
large momentum transfer, perturbative-QCD methods may be used. 
As was discussed a long time ago~\cite{Owens:1986mp}, 
in perturbative high-$p_T$ processes there are two production mechanisms for photons. 
The photon may be directly produced in the hard scattering process through its ``pointlike'' QED coupling 
to a quark. Such contributions are usually referred to as ``direct". On the other hand, photons may also 
emerge in jet fragmentation, when a quark, antiquark, or gluon emerging from a QCD 
hard-scattering process fragments into a photon plus a number of hadrons. The need for introducing 
such a ``photon fragmentation'' contribution is physically motivated by the fact that the photon may result,
for example, from conversion of a high-$p_T$ $\rho$ meson.
Furthermore, at higher orders, the perturbative direct component contains divergencies from 
configurations where the photon and a final-state quark become collinear. These are long-distance 
contributions that naturally signify the need for non-perturbative fragmentation functions 
into which they can be absorbed.

Using the single-inclusive process $pp\to \gamma X$ as an example, QCD perturbation
theory thus predicts the following schematic structure of the factorized cross section:
\beq
d\sigma = \sum_{a,b}
f_a\otimes f_b\otimes \bigg[d\hat{\sigma}_{ab}^\gamma+
\sum_c d \hat{\sigma}_{ab}^c \otimes D_c^\gamma\bigg]\,.
\eeq
Here the sums run over all partons (quarks, antiquarks and gluons). 
The part involving the sum over partons $c$ is the fragmentation component, 
while the other part is the direct one. Each part has its own partonic hard-scattering cross 
sections for producing either directly a photon, or a parton $c$ that subsequently fragments
into the photon as described by the fragmentation functions $D_c^\gamma\equiv
D_c^\gamma(z,\mu)$, where $z$ is the fraction of the parton's momentum picked
up by the photon and $\mu$ the scale at which the fragmentation function is probed.
Both the direct and the fragmentation parts involve parton distribution functions
in the proton, denoted by $f_a$ and $f_b$. The symbols $\otimes$ represent
appropriate integrations over momentum fractions of the various participating
partons. 

Theoretical calculations of photon production cross sections in hadronic
collisions thus rely on three ingredients: 
\begin{enumerate}
\item[(i)] knowledge of parton distributions, 
\item[(ii)] precision calculations of the partonic cross sections, and 
\item[(iii)] parton-to-photon fragmentation functions. 
\end{enumerate}
The parton distribution functions of protons are by now 
very well known. A lot of work has gone into (ii) over the past three decades.
For single-inclusive photon production, full next-to-leading order (NLO) computations
of the partonic cross sections for the direct~\cite{singlephoton} as well
as for the fragmentation part~\cite{singleparton} have been performed, and even
all-order resummations of large logarithmic threshold corrections are available for
both parts~\cite{resu}.

Rather little is known, on the other hand, about the photon fragmentation functions. 
Some information is available from measurements at LEP~\cite{Buskulic:1995au}, but
with rather large uncertainties. Theo\-retical 
predictions~\cite{Bourhis:1997yu,Gluck:1992zx,Aurenche:1992yc,GehrmannDeRidder:1997wx}  
for the photon fragmentation functions are compatible with these sparse data. Based on such sets of $D_c^\gamma$,
one estimates that fragmentation photons contribute about 10-30\% to the cross section for $pp\to \gamma X$
in the fixed-target regime~\cite{Vogelsang:1995bg}. At high-energy colliders, the fragmentation 
contribution is typically predicted to be only 10\% or less~\cite{Gluck:1994iz}. The reason for this is that 
at colliders one introduces so-called isolation cuts. To isolate a photon candidate, 
one basically demands that the hadronic energy in a cone around the photon be limited to a certain value.
Isolation suppresses the large background from $\pi^0\to \gamma\gamma$ decay and also 
reduces the fragmentation contribution~\cite{Gluck:1994iz}
since it basically confines it to very high values of the fragmentation variable $z$. 

At any rate, the size of the fragmentation contribution arguably presents the largest uncertainty in predictions 
of photon production cross sections. In the present paper, we propose a new promising method to experimentally 
access the $D_c^\gamma$ in hadronic collisions. The idea is to identify a photon as part of a fully reconstructed 
jet. One measures the transverse momenta of the photon and the jet, $p_T$ and $p_T^\jet$, respectively. 
As we shall show, their ratio,
\beq\label{defzgamma}
z_\gamma\,\equiv\,\frac{p_T}{p_T^\jet}\,,
\eeq
 offers direct scans of the fragmentation functions $D_c^\gamma(z,\mu)$ in $z$. This is an advantage
            over the inclusive cross section $pp\to \gamma X$, which may also provide information on the photon 
            fragmentation functions~\cite{Klasen:2014xfa} but typically samples them over a broad region in $z$. We develop
            the formalism for the ``same-side'' photon-jet observable to NLO accuracy. Using the approximation of relatively narrow 
reconstructed jets, we are able to derive analytical results for the relevant NLO cross sections. In our previous 
publication~\cite{Kaufmann:2015hma} (see also~\cite{Arleo:2013tya,Chien:2015ctp,Procura:2011aq} for related work), we have already presented the 
corresponding NLO calculation for the process $pp\to(\jet\,h)\,X$, where $h$ denotes an identified hadron inside the jet, 
and demonstrated that this process opens a promising window on the determination of the parton-to-hadron fragmentation 
functions $D_c^h$. Evidently, the background from $\pi^0\to\gamma\gamma$ decay mentioned above will also be relevant 
in our case and hence will also be analyzed in our paper.

We stress that the observable we have in mind is different from the ``away-side'' photon-jet correlations 
considered in Ref.~\cite{Belghobsi:2009hx} and provides a kinematically simpler and more direct access to the 
$D_c^\gamma$. 
In particular, in our case it is natural to divide the cross section for $pp \to (\jet \gamma)X$
             by the single-inclusive jet one for $pp \to \jet X$, in which case many theoretical uncertainties 
             related to the choice of parton distributions or (initial-state) factorization scale will cancel out.
We also note that same-side photon-jet observables have been studied in the contexts of $e^+e^-$ 
annihilation \cite{Glover:1993xc,GehrmannDeRidder:1997wx} 
and $ep$ collisions \cite{GehrmannDeRidder:2006vn}, where they were shown to provide
access to the $D_q^\gamma$. However, the accuracy achievable in experiment is rather limited
for these reactions, as the LEP results~\cite{Buskulic:1995au} show. More importantly, a drawback of these processes 
is that the gluon-to-photon fragmentation function can be probed only indirectly by evolution or higher-order corrections. 
This is different for the process $pp\to(\jet\,\gamma)\,X$ which probes all photon fragmentation functions
in the same hadronic environment as relevant for photon signals at colliders.

Our paper is organized as follows. In Sec.~\ref{sec2} we present our NLO calculation of $pp \to (\jet \gamma)X$.
We shall see that the results can be formulated in terms of simple photonic ``jet'' functions. Section~\ref{sec3}
describes calculations of the pion decay background, taking into account basic experimental considerations
concerning pion reconstruction. In Sec.~\ref{sec4} we present phenomenological results for 
$pp \to (\jet \gamma)X$ relevant for collisions at RHIC and the LHC.

%%%%%%%%%%%%%%%%%%%%%%%%%%%%%%%%%%%%%%%%%%%%%%%%%%%%%%
%%%%%%%%%%%%%%%%%%%%%%%%%%%%%%%%%%%%%%%%%%%%%%%%%%%%%%
\section{Same-side jet-plus-photon production at NLO\label{sec2}}

We consider a high-$p_T$ photon produced inside a jet. 
Our approach will be to consider the photon as part of the jet. To this
end, we follow Ref.~\cite{Glover:1993xc,GehrmannDeRidder:1997wx} to cluster hadrons and photons ``democratically'' 
into mixed hadronic and electromagnetic 
jets, a procedure that may be applied to any of the commonly used jet algorithms~\cite{algo:cone,algo:kt,algo:JET}. 
In terms of a perturbative-QCD calculation it means that the photon is included as an additional parton.

At lowest order in QCD, for $z_\gamma<1$, there are only fragmentation contributions to the cross section for 
$pp\to(\jet\,\gamma)\,X$. It is this feature that makes the cross section a powerful probe of the 
photon fragmentation functions.
Our strategy for performing the NLO calculation is based on our previous work~\cite{Kaufmann:2015hma} on the process 
$pp\to(\jet\,h)\,X$. A key element of that study was the ``narrow jet approximation (NJA)''~\cite{Jager:2004jh} 
(see also~\cite{Dasgupta:2016bnd}), 
which assumes that the parameter $\mathcal{R}$ specifying the jet ``opening'' is relatively small. 
This allows for an analytical calculation 
of the NLO partonic cross sections. The NJA is accurate up to corrections of order ${\cal R}^2$. As was shown in 
Refs.~\cite{Jager:2004jh}, this is sufficient for essentially all jet opening parameters used in experiment. 

In the language of Ref.~\cite{Kaufmann:2015hma}, the NLO factorized cross section for 
$pp\to(\jet\,\gamma)\,X$ may be written as
%\begin{widetext}
\beeq\label{Xsection3}
\frac{d\sigma^{pp\rightarrow (\jet\,\gamma)X}}{dp_T^{\jet} d\eta^\jet d z_\gamma} &= &
\sum_{\begin{subarray}{c} a,b,c\,\in\,\\ \{q,\bar{q},g,\gamma\}\end{subarray}} \int_{x_a^{\mathrm{min}}}^1 dx_a\, f_a(x_a,\mu_F)
\int_{x_b^{\mathrm{min}}}^1 dx_b\, f_b(x_b,\mu_F)\,
\int_{z_c^{\mathrm{min}}}^1 dz_c\,
\frac{d\hat{\sigma}_{ab}^c(\hat{s},\hat{p}_T,\hat\eta,\mu_F,\mu_F^\prime,\mu_R)}{dp_T^{\jet} d\eta^\jet}\nn\\[2mm]
&\times&\sum_{\begin{subarray}{c}d\,\in\, \\ \{q,\bar{q},g,\gamma\}\end{subarray}} 
j_{c\to d}\left(z_c,\frac{{\cal R}\,p_T^{\mathrm{jet}}}{\mu_F^\prime}\right)
\sum_{\begin{subarray}{c} c'\,\in\,\\ \{q,\bar{q},g,\gamma\}\end{subarray}}\int_{z_\gamma}^1 \frac{dz_p}{z_p}\,
\tilde{j}_{d\to c'}\left(z_p,\frac{{\cal R}\,p_T^{\mathrm{jet}}}{\mu_F^{\prime\prime}}\right)
\mathcal{D}_{c'}^\gamma\left(\frac{z_\gamma}{z_p},\mu_F^{\prime\prime}\right).
\eeeq
%\end{widetext}
Apart from the variable $z_\gamma$ defined in~(\ref{defzgamma}) the cross section is specified by the jet's 
transverse momentum $p_T^\jet$ and rapidity $\eta^\jet$, the latter being equal to the 
photon rapidity $\eta$ in the NJA. In Eq.~(\ref{Xsection3}) we have furthermore defined
\beeq\label{xmin}
&& x_a^{\mathrm{min}}\,=\,\frac{p_T^{\mathrm{jet}}{\mathrm{e}}^{\eta^{\mathrm{jet}}}}
{\sqrt{S}-p_T^{\mathrm{jet}}{\mathrm{e}}^{-\eta^{\mathrm{jet}}}}
\,,\;\;\; x_b^{\mathrm{min}}\,=\,\frac{x_a p_T^{\mathrm{jet}}{\mathrm{e}}^{-\eta^{\mathrm{jet}}}}
{x_a\sqrt{S}-p_T^{\mathrm{jet}}{\mathrm{e}}^{\eta^{\mathrm{jet}}}}\,,\nn\\[2mm]
&& z_c^{\mathrm{min}}\,=\,\frac{p_T^{\mathrm{jet}}}{\sqrt{S}}\left(\frac{{{\mathrm{e}}^{\eta^{\mathrm{jet}}}}}{x_a}+
\frac{{{\mathrm{e}}^{-\eta^{\mathrm{jet}}}}}{x_b}\right)\,,
\eeeq
where $\sqrt{S}$ is the hadronic c.m.s. energy. The various partonic variables are 
$\hat{s}=x_ax_bS$, $\hat{p}_T=p_T^{\mathrm{jet}}/z_c$, 
$\hat{\eta} = \eta^\jet - \frac{1}{2} \log(x_a/x_b)$. 

Equation~(\ref{Xsection3}) has been written in such a way that it includes both the 
direct and the fragmentation contributions. This is achieved by introducing the function
\beq\label{defD}
\mathcal{D}_{c'}^\gamma(z,\mu) \equiv \delta(1-z) \delta_{c'\gamma} + D_{c'}^{\gamma}(z,\mu) \,
(1-\delta_{c'\gamma})\,,
\eeq
where the first part corresponds to the direct contribution and the second part to the fragmentation one. 
The partonic cross sections $d\hat{\sigma}_{ab}^c$
(with $c=\gamma,q,\bar{q},g$) in~(\ref{Xsection3}) are the usual NLO {\it single-inclusive photon/parton cross 
sections} as computed in Refs.~\cite{singlephoton} and~\cite{singleparton}, respectively. Their perturbative 
expansions to NLO, but to lowest order in the electromagnetic coupling $\alpha$, read
\beeq\label{partXsecExp}
d\hat{\sigma}_{ab}^\gamma &=& \alpha\as\left[d\hat{\sigma}_{ab}^{\gamma,(0)} + \frac{\as}{\pi}
d\hat{\sigma}_{ab}^{\gamma,(1)} + \mathcal{O}(\as^2)  \right]  \,,\nn\\[2mm]
d\hat{\sigma}_{ab}^{c\neq\gamma} &=& \alpha_s^2 \left[d\hat{\sigma}_{ab}^{c,(0)} + 
\frac{\as}{\pi}d\hat{\sigma}_{ab}^{c,(1)} + \mathcal{O}(\as^2)  \right] \,.
\eeeq
Evidently, in order to compute the cross section for production of photons inside a jet,
one needs corresponding partonic cross sections for this observable. These necessarily
differ from the single-inclusive cross sections $d\hat{\sigma}_{ab}^\gamma,d\hat{\sigma}_{ab}^c$.
As was shown in Ref.~\cite{Kaufmann:2015hma}, it is relatively straightforward to convert
the $d\hat{\sigma}_{ab}^\gamma,d\hat{\sigma}_{ab}^c$ into the cross sections that we need for 
same-side photon (or hadron) plus jet production. This is achieved by introducing suitable perturbative
jet functions $j_{c\to d}$ and $\tilde{j}_{d\to c'}$. The former describe the formation of a jet ``consisting'' 
of parton $d$ which has taken the fraction $z_c$ of the parent parton's momentum, while the latter 
represent the ``partonic fragmentation'' of parton $d$ to parton $c'$ inside the jet, with momentum 
fraction $z_p$. 

Before discussing the jet functions in more detail, we note that we have introduced several scales
in Eq.~(\ref{Xsection3}): The renormalization scale $\mu_R$, the initial-state factorization scale 
$\mu_F$ and two final-state factorization scales $\mu_F^\prime$ and $\mu_F^{\prime\prime}$, respectively. 
Of the latter two, $\mu_F^\prime$ is an artifact of the way we organize our calculation, formulating 
the cross section in terms of {\it single-inclusive} parton cross sections. $\mu_F^\prime$ was 
originally introduced in the mass factorization procedure for the $d\hat{\sigma}_{ab}^c$. 
Now, however, it cancels exactly between the $d\hat{\sigma}_{ab}^c$ and the jet functions $j_{c\to d}$
due to the fact that we foremost define our observable by requiring an infrared- and collinear safe jet 
in the final state. Hence, $\mu_F^\prime$ is not present in the final result. 
The actual final-state factorization scale which survives in our  final answer is $\mu_F^{\prime\prime}$. 
This scale appears because we are asking for an observed photon in the final state as part of the jet. 
For details, see Ref.~\cite{Kaufmann:2015hma}.

Throughout this paper (and as anticipated in Eq.~(\ref{partXsecExp})), we will always
work to lowest order ${\cal O}(\alpha)$ in the electromagnetic coupling $\alpha$. This means that the
initial partons $a,b$ in Eq.~(\ref{Xsection3}) will never be photons in our calculations 
since the requirement that there also is a photon in the final state would make the
corresponding contribution to the cross section $\mathcal{O}(\alpha^2)$. This holds true
even for the photon fragmentation part since the fragmentation functions $D_c^\gamma$ 
always provide an additional power of $\alpha$. In order to specify the jet functions, it is
therefore sufficient to just consider a fixed initial partonic state $ab$ ($a,b\neq\gamma$) 
in Eq.~\eqref{Xsection3} and disregard the parton distribution functions. Schematically, 
we then have the expression 
\beq\label{expr}
\sum_{c,d,c^\prime}d\hat{\sigma}_{ab}^c \otimes\, j_{c\to d} \otimes \tilde{j}_{d\to c'} 
\otimes \mathcal{D}_{c'}^\gamma\,.
\eeq
The labels $c,d,c^\prime$ run over all QCD partons, but can also represent a photon. 
The jet functions with both $c\neq \gamma$ and $c'\neq \gamma$ contribute only to the
fragmentation part. They are identical to those appearing in our earlier calculation~\cite{Kaufmann:2015hma}
for hadron production inside jets, where they were given to NLO. To LO, they are all diagonal, $j_{q\to q}=j_{g\to g}=\delta(1-z_c)$ 
and $\tilde{j}_{q\to q}=\tilde{j}_{g\to g}=\delta(1-z_p)$. This shows that at this order the cross section for $pp\to(\jet\,\gamma)\,X$ 
directly probes the photon fragmentation functions at value $z=z_\gamma$. 

Starting at NLO, there is also a direct component of the cross section for photon-in-jet
production. The corresponding contributions are always associated with either $c=\gamma$ or 
$c'=\gamma$. Being direct contributions, they always come with the piece $\propto \delta(1-z)$ in 
$\mathcal{D}_{c^\prime}^\gamma$ in~(\ref{defD}). We now discuss the various possibilities at ${\cal O}(\alpha)$:\\[1mm]
{\bf (i)} Case $c=c'=\gamma$: Here the expression in~(\ref{expr}) becomes
\beq
\sum_{d}\underbrace{d\hat{\sigma}_{ab}^\gamma}_{\mathcal{O}(\alpha\as)} \otimes \,j_{\gamma\to d} 
\otimes \tilde{j}_{d\to \gamma} \otimes \underbrace{\mathcal{D}_{\gamma}^\gamma}_{\mathcal{O}(1)}\,.
\eeq
As indicated, the partonic cross sections $d\hat{\sigma}_{ab}^\gamma$ start at order $\mathcal{O}(\alpha\as)$ 
(see Eq.~\eqref{partXsecExp}). Since all jet functions for $d\neq\gamma$ are of order $\mathcal{O}(\alpha)$, 
they would give rise to higher-order contributions in $\alpha$. Hence, only $d=\gamma$ and
$j_{\gamma\to\gamma}=\delta(1-z_c)$ and $\tilde{j}_{\gamma\to\gamma}=\delta(1-z_p)$ is possible at
${\cal O}(\alpha)$. Thus, this part gives precisely the direct part of the full NLO single-inclusive photon production 
cross section. However, because of the $\delta$-functions from $\tilde{j}_{\gamma\to\gamma}$ and
$\mathcal{D}_{\gamma}^\gamma$, the entire piece comes 
with a factor $\delta(1-z_\gamma)$, as is easily seen by insertion into Eq.~(\ref{Xsection3}). This in turn
implies that for any measurement of the cross section carried out at $z_\gamma < 1$ this part of the 
cross section will not contribute.\footnote{We note that the region $z_\gamma\to1$ could be interesting 
when isolation cuts are imposed on the photon. In this region we have a photon carrying most of the 
jet's energy and accompanied by very little hadronic energy. For such studies at very high $z_\gamma$
it may be preferred to integrate over a range of $z_\gamma$ that includes the endpoint at $z_\gamma=1$,
in which case the contributions $\propto \delta(1-z_\gamma)$ will need to be kept.}

In cases {\bf (ii)} and {\bf (iii)} we have $c=q$ and $c'=\gamma$ so that the expression in~(\ref{expr}) becomes
\beq
\sum_{d} \underbrace{d\hat{\sigma}_{ab}^q}_{\mathcal{O}(\as^2)} \otimes \,j_{q\to d} \otimes 
\tilde{j}_{d\to \gamma} \otimes \underbrace{\mathcal{D}_{\gamma}^\gamma}_{\mathcal{O}(1)}\,.
\eeq
To lowest order in $\alpha$, we can then either have $d=\gamma$ or $d=q$ in Eq.~(\ref{Xsection3}). Thus 
we further distinguish: \\[1mm]
{\bf (ii)} Case $c=q,c'=\gamma,d=\gamma$: We have $\tilde{j}_{\gamma\to\gamma}=\delta(1-z_p)$ and, 
in the $\overline{{\mathrm{MS}}}$ scheme,
\beq
j_{q\to\gamma}(z,\lambda)=-\frac{\alpha}{2\pi}e_q^2
\left[P_{\gamma q}(z)\log\left( \lambda^2(1-z)^2\right)+ 
z\right],
\eeq
where $e_q$ is the quark's fractional charge, and where
\beq\label{Pgq}
P_{\gamma q}(z) = \frac{1+(1-z)^2}{z}\,.
\eeq
This contribution is again proportional to  $\delta(1-z_\gamma)$ and does not contribute to measurements at $z_\gamma < 1$. 
However, this becomes different in \\[1mm]
{\bf (iii)} case $c=q,c'=\gamma,d=q$, where $j_{q\to q}=\delta(1-z_c)$ and, 
in the $\overline{{\mathrm{MS}}}$ scheme and for the anti-$k_t$ jet algorithm:
\beq\label{MSb}
\tilde{j}_{q\to\gamma}^{\overline{{\mathrm{MS}}}}(z,\lambda) = \frac{\alpha}{2\pi} e_q^2\,\left[ P_{\gamma q}(z)
\log\left(\lambda^2 z^2 (1-z)^2 \right)+ z \right].
\eeq
The contribution associated with this jet function is, at this order, the {\it only} direct part
of the cross section that is present at $z_\gamma<1$. It is genuinely NLO, and its
logarithmic dependence on the scale $\mu_F^{\prime\prime}$ correctly compensates 
the scale dependence associated with the inhomogeneous 
piece~\cite{Bourhis:1997yu,Gluck:1992zx,Aurenche:1992yc,GehrmannDeRidder:1997wx} 
in the evolution of the $D_c^\gamma$.

The final possibility is \\[1mm]
{\bf (iv)} case $c=g,c'=\gamma$: Here we need to `convert' a gluon exiting the hard scattering to an observed
photon. This may happen through the combination $j_{g \to q}\, \tilde{j}_{q \to \gamma}$. The product of these
functions is $\mathcal{O}(\alpha \alpha_s)$, which, when combined with the $\alpha_s^2 $ from
the hard process is an order too high in $\alpha_s$, that is, beyond NLO. The only contribution
for $c=g$ arises when the photon is produced in fragmentation, as discussed above.

We collect all jet functions in the Appendix, both for the fragmentation and for the direct cases, 
and for various types of jet algorithms. From the above discussion, 
we conclude that as long as we stay at $z_\gamma<1$ case {\bf (iii)} 
generates the only non-vanishing direct contributions at ${\cal O}(\alpha)$ and NLO. 
We observe that the corresponding jet function in Eq.~(\ref{MSb}) shows a logarithmic
divergence at $z\to 1$. A behavior of this type is quite familiar for NLO direct (pointlike) contributions
in hadronic processes involving photons, where it is also exhibited by the corresponding photonic
coefficient functions for space-like or time-like photon structure functions~\cite{Gluck:1991ee,Gluck:1992zx}. 
This can become an issue, in principle, due to the fact that the coefficient functions appear
directly in the structure function, potentially creating an unphysical divergent 
behavior of the structure function at high Bjorken-$x$. However, as was shown in 
Refs.~\cite{Gluck:1991ee,Gluck:1992zx}, for photon structure functions the situation may
be remedied by adopting a factorization scheme different from the $\overline{{\mathrm{MS}}}$ scheme
for the pointlike contribution. This scheme is known as DIS$_\gamma$ scheme and will also be
our choice in this paper as it leads to an overall better perturbative stability. Using the results
of~\cite{Gluck:1992zx} the scheme transformation
is easily performed, and we find for the jet function in the DIS$_\gamma$ scheme:
\beeq\label{disg}
\tilde{j}^{\mathrm{DIS}_\gamma}_{q\to\gamma}(z,\lambda) &\equiv& \tilde{j}^{\overline{{\mathrm{MS}}}}_{q\to\gamma}(z,\lambda) - 
\frac{\alpha}{2\pi} e_q^2 \left[ P_{\gamma q}(z) \log\left(z^2(1-z) \right) - 2\frac{1-z}{z} \right]\nn\\[2mm]
&=& \frac{\alpha}{2\pi} e_q^2\,P_{\gamma q}(z)\left[
\log\left(\lambda^2(1-z)\right)+ 1\right].
\eeeq
All other jet functions remain unaffected by the transformation. 
Interestingly, we find that even in the DIS$_\gamma$ scheme
a logarithmic term $\propto \log(1-z)$ remains in the jet function. The reason for this can be traced
back to the fact that, compared to the inclusive structure function $F_1^\gamma$, the phase space 
for our photon-in-jet observable is different, being less inclusive and hence 
giving rise to two factors of $\propto \log(1-z)$ in the
$\overline{{\mathrm{MS}}}$ scheme, only one of which is subtracted when transforming to the 
DIS$_\gamma$ scheme. In any case, the remaining logarithm in~(\ref{disg}) may be eliminated
by a scale choice $(\mu_F^{\prime\prime})^2\propto (1-z_\gamma)$. 
As mentioned, the jet function in~(\ref{disg}) refers to the anti-$k_t$ jet algorithm. 
The results for other algorithms may be found in the Appendix for the $\overline{{\mathrm{MS}}}$ scheme
and may then be transformed to the DIS$_\gamma$ scheme in the same fashion as above. 

For later reference, we finally note that the cross section \eqref{Xsection3} may also be expressed in terms of 
photon kinematics,
\beq\label{Xsec:photonkin}
\frac{d\sigma^{pp\rightarrow (\jet\,\gamma)X}}{dp_T d\eta d z_\gamma}\left(p_T,\eta,z_\gamma\right)= \frac{1}{z_\gamma} 
\frac{d\sigma^{pp\rightarrow (\jet\,\gamma)X}}{dp_T^{\jet} d\eta^\jet d z_\gamma} 
\left(p_T^{\jet} = \frac{p_T}{z_\gamma},\eta,z_\gamma \right)\,.
\eeq

%%%%%%%%%%%%%%%%%%%%%%%%%%%%%%%%%%%%%%%%%%%%%%%%%%%%%%
%%%%%%%%%%%%%%%%%%%%%%%%%%%%%%%%%%%%%%%%%%%%%%%%%%%%%%
\section{Pion decay background \label{sec:pions}\label{sec3}}

Production of $\pi^0$ with the subsequent decay $\pi^0 \to \gamma\gamma$ 
(with a branching ratio of $\approx 98.8\%$ \cite{Agashe:2014kda})
provides the largest source of unwanted background photons. There are two main components
of this background~\cite{Adare:2012yt,carl}. First, at high pion transverse momentum, the two decay photons can become
almost collinear and may not be resolved as two separate photons but rather as a single electromagnetic 
signal. We refer to this part of the background as ``B1''. 
Secondly, even when the two decay photons are well separated in angle, the decay may be very
asymmetric in energy. It may happen that one of the decay photons is relatively soft, so that it 
is below the threshold for detection. In this case, only the high-energy decay photon will be seen
and misinterpreted as a single prompt photon. This part of the background, denoted as ``B2'',
is relevant toward lower transverse momentum, where the pion production cross section is large. 

Using our previous calculation~\cite{Kaufmann:2015hma}, it is relatively straightforward to estimate 
the sizes of B1 and B2. From the axial anomaly prediction \cite{WZW} in the chiral limit we have a constant
decay matrix element, so that the normalized differential decay rate follows just from phase space considerations:
\beq
\frac{1}{\Gamma} \frac{d\Gamma}{dE_1} =  \frac{1}{|\vec{P}_\pi|}\,,
\eeq
where $E_1$ is the energy of one of the decay photons and $\vec{P}_\pi$ denotes the pion three-momentum 
in the hadronic c.m.s. As we are interested in the angular separation of the two photons, 
we relate $E_1$ to $\cos\theta_{12}$, where $\theta_{12}$ is the angle between the momenta of the 
two decay photons:
\beq\label{costheta12}
\cos\theta_{12} = 1-\frac{m_\pi^2}{2E_1(E_\pi-E_1)}\,.
\eeq
We then obtain
\beq
\frac{1}{\Gamma} \frac{d\Gamma}{d\cos\theta_{12}} = \frac{m_\pi^2}{|\vec{P}_\pi|}\frac{1}{(1-\cos\theta_{12})^2
\sqrt{E_\pi^2 - \frac{2m_\pi^2}{1-\cos\theta_{12}}}} \,, \label{gammafinal}
\eeq
where $E_\pi$ is the pion energy. For B1 we only need to impose the condition 
that $\theta_{12}$ be smaller than the angle the detector can resolve, so that the photons will be seen as a 
single electromagnetic signal. This happens when
\beq\label{BG}
1-\cos\theta_{12} \le \frac{\delta\phi^2 + \delta\eta^2}{2\cosh^2\eta}\,\equiv\,
 \frac{\rho^2}{2\cosh^2\eta}\,,
\eeq
where $\delta \phi$ and $\delta \eta$ are the detector resolutions in azimuthal angle and rapidity,
respectively. To estimate B1 numerically, we simply have to integrate the normalized 
decay spectrum~(\ref{gammafinal}) subject to the constraint~(\ref{BG}), i.e.
\beeq
\int_{1 - \rho^2/(2\cosh^2\eta)}^{1-2(1-\beta^2)} d\cos\theta_{12}\, \frac{1}{\Gamma} \frac{d\Gamma}{d\cos\theta_{12}}
= \frac{1}{|\vec{P}_\pi|}\sqrt{E_\pi^2 - 4m_\pi^2 \frac{\cosh^2\eta}{\rho^2}}\, 
\Theta\left(E_\pi^2 - 4m_\pi^2 \frac{\cosh^2\eta}{\rho^2}\right)\,,\label{br}
\eeeq
where $\Theta(x)$ denotes the Heaviside function, and multiply the result by the 
cross section for $pp\to(\jet\,\pi^0)\,X$ from our previous calculation~\cite{Kaufmann:2015hma}. Here we have introduced the
Lorentz factor $\beta = |\vec{P}_\pi| / E_\pi$.

For B2, for which one of the two decay photons is not detected and the other photon has transverse
momentum $p_T$, one needs to take into account that the decaying pion must have had a correspondingly 
higher transverse momentum or energy. This gives rise to a convoluted structure of the cross section
for this part of the background. Denoting the energy threshold for photon detection by $\ep$, 
we have
\beq\label{B2def}
\frac{d\sigma^{\mathrm{B2}}}{d\eta^\jet dp_T^\gamma dz_\gamma} = \int_{p_T^\gamma}^\infty 
\frac{dp_T^\pi}{p_T^\pi}\,\frac{d\sigma^{(\jet\,\pi^0)}}{d\eta^\jet dp_T^\pi dz_\pi} \frac{1}{\Gamma} 
\frac{d\Gamma (x=p_T^\gamma / p_T^\pi)}{dx}\,,
\eeq
where
\beq\label{def:dGammadx}
\frac{1}{\Gamma} \frac{d\Gamma (x)}{dx} = \frac{2}{\beta} \Theta\left[\frac{1+\beta}{2}-x\right] 
\Theta\left[x-\max \left(1-\frac{\ep}{E_\pi}, \frac{1}{2}\left(1+\sqrt{1-\frac{4 m_\pi^2 \cosh^2\eta}{\rho^2 E_\pi^2}}\right)\right)\right]\,.
\eeq
We have introduced the variable $x$ as the energy fraction of one photon with respect to the pion energy, 
$E_1 = x E_\pi$, and hence $E_2 = (1-x) E_\pi$.
Furthermore, $p_T^\pi$ is the transverse momentum of the decaying pion and $z_\pi \equiv p_T^\pi / \hat{p}_T^\jet$ 
is defined in analogy to Eq.~\eqref{defzgamma}, with $\hat{p}_T^\jet$ the transverse momentum of the jet containing the 
pion. Clearly, we have $E_\pi = E_\gamma + E_2$ and hence $p_T^\gamma = x p_T^\pi$, since we assume that 
the other photon's energy, $E_2$, is not seen by the detector as it falls below the detector's energy threshold. 
A further consequence of this is that the photon with energy $E_2$ is not counted toward the jet energy.
As a result, we have $p_T^\jet = x \hat{p}_T^\jet$. Combining these relations, we find that $z_\pi = z_\gamma$.

We now rewrite the integration in Eq.~\eqref{B2def} in terms of the energy fraction $x$ and use 
Eq.~\eqref{Xsec:photonkin} to express everything in terms of the jet kinematics. Using $z_\pi=z_\gamma$ we finally arrive at
\beq
\frac{d\sigma^{\mathrm{B2}}}{dp_T^{\mathrm{jet}} d\eta^{\mathrm{jet}}dz_\gamma}
\,=\,\int_0^1 dx\,\frac{d\sigma^{({\mathrm{jet}}\,\pi^0)}}{d\hat{p}_T^{\mathrm{jet}} d\eta^{\mathrm{jet}}dz_\pi}\,
\frac{1}{\Gamma}\frac{d\Gamma(x)}{dx}\,.
\eeq
The integration limits in this equation are implemented implicitly in the definition 
of $\frac{1}{\Gamma} \frac{d\Gamma (x)}{dx}$ via Heaviside functions, 
see Eq.~\eqref{def:dGammadx}. They follow from the conditions that the decay be kinematically possible,
that the unobserved decay photon must have energy below $\ep$, and that the two photons have an 
opening angle larger than the detector resolution. Explicitly, one has 
\beq
1+\frac{m_\pi^2}{4E_\gamma^2}\leq x^{-1}\leq 
\min\left(1+\frac{\ep}{E_\gamma},1+\frac{m_\pi^2}{\rho^2 p_T^2}\right)\,,
\eeq
with $E_\gamma=p_T\cosh\eta$ the observed photon's energy.

\begin{figure}[!ht]
	\centering
	\includegraphics[width=0.9\textwidth]{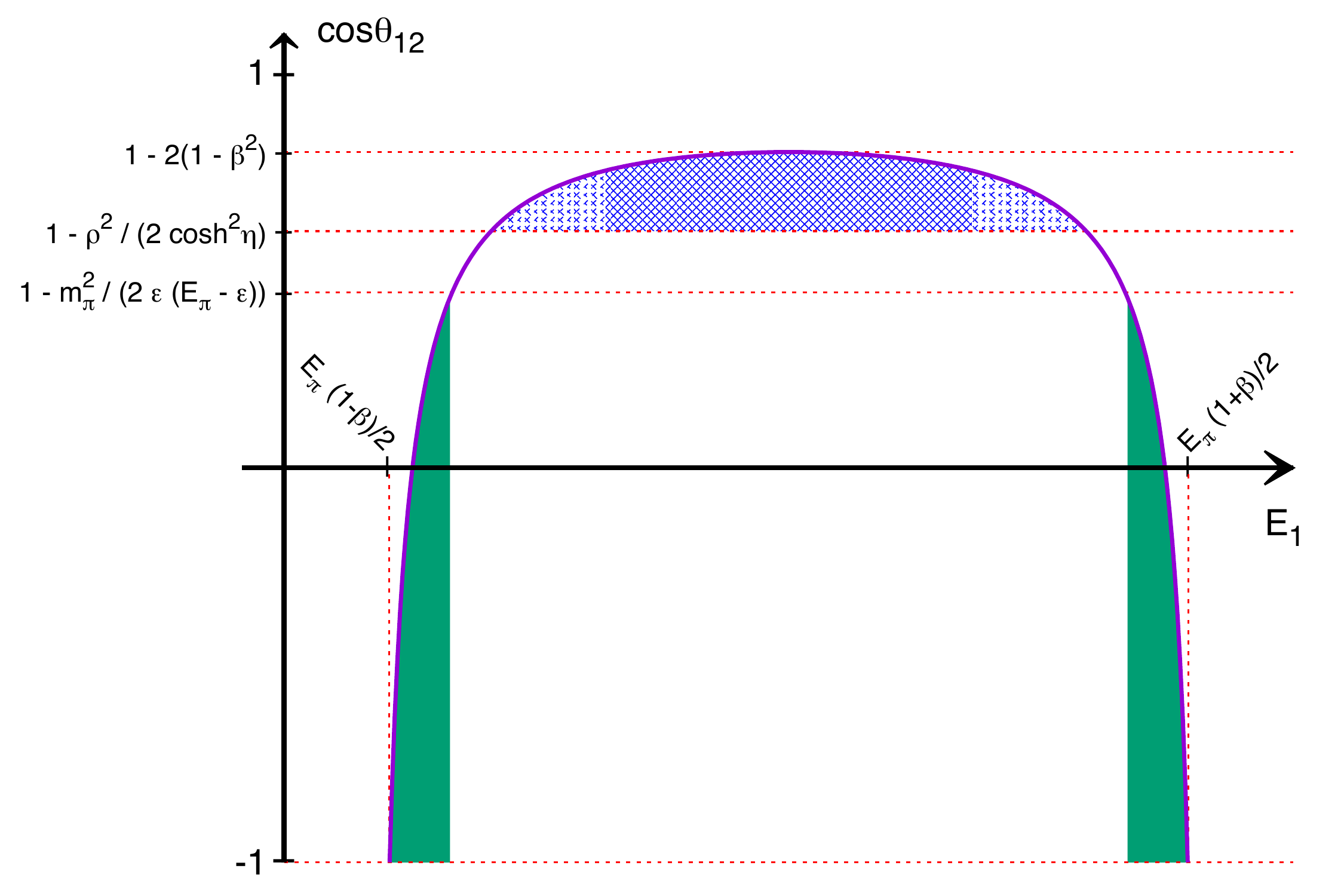}
	%\vspace*{-3.5cm}
	\caption{{\sl The cosine of the angle between the two photons, $\cos\theta_{12}$, as a function of the
	                    photon energy $E_1$, according to Eq.~\eqref{costheta12}. 
	                    The shaded areas correspond to B2 while the hatched one describes B1.}}
	\label{cost12}
\end{figure}

To obtain a better understanding of the phase space for the two backgrounds and how they are connected, it is
instructive to plot $\cos\theta_{12}$ as a function of $E_1$ according to Eq.~\eqref{costheta12}.
This is done in Fig.~\ref{cost12}. Without any restrictions, the region underneath the solid line 
would be the available phase space for the two photon decay. If the pion could be reconstructed
experimentally in this whole region, there would be no pion decay background. As described above, 
the experimental limitations lead to some regions where only one electromagnetic signal is seen.
In the upper hatched region representing B1 the two photons are close together and cannot be seen
as separate photons by the detector. In the shaded regions (B2) left and right, they are well separated
in angle, but one of them has energy below threshold and is not seen either. 
The full background comes from the sum of these regions. Only in the the remaining part can
both photons be seen and the parent pion be reconstructed. We note that Fig.~\ref{cost12} shows an 
idealized picture where we have chosen the pion energy and the parameters $\rho$ and $\varepsilon$ 
such that the main qualitative effects are well visible. If we were to use more realistic values, the 
plot would look close to a rectangular function, with $\cos\theta_{12}$ very closely reaching
unity at the upper end. The hatched and shaded areas would shrink in size, although they still
make numerically very sizable contributions as we shall see in the next section. We also note
that, depending on the kinematics, the areas of B1 and B2 may overlap. Hence, in the numerical calculation 
one has to carefully avoid double counting. 

We finally emphasize that our calculations of the pion backgrounds are not limited to photon-in-jet
production, but can also be applied to any other photon observable. Of course, one has to combine the 
decay results with the appropriate pion cross sections corresponding to the photon
observable one is interested in. For example, to obtain a background estimation for single-inclusive photon 
production one would use a single-inclusive pion cross section and apply the expressions for B1 and B2.

%%%%%%%%%%%%%%%%%%%%%%%%%%%%%%%%%%%%%%%%%%%%%%%%%%%%%%
%%%%%%%%%%%%%%%%%%%%%%%%%%%%%%%%%%%%%%%%%%%%%%%%%%%%%%
\section{Phenomenological results \label{pheno} \label{sec4}}

\begin{figure}[t!]
	\centering
	\includegraphics[width=0.9\textwidth]{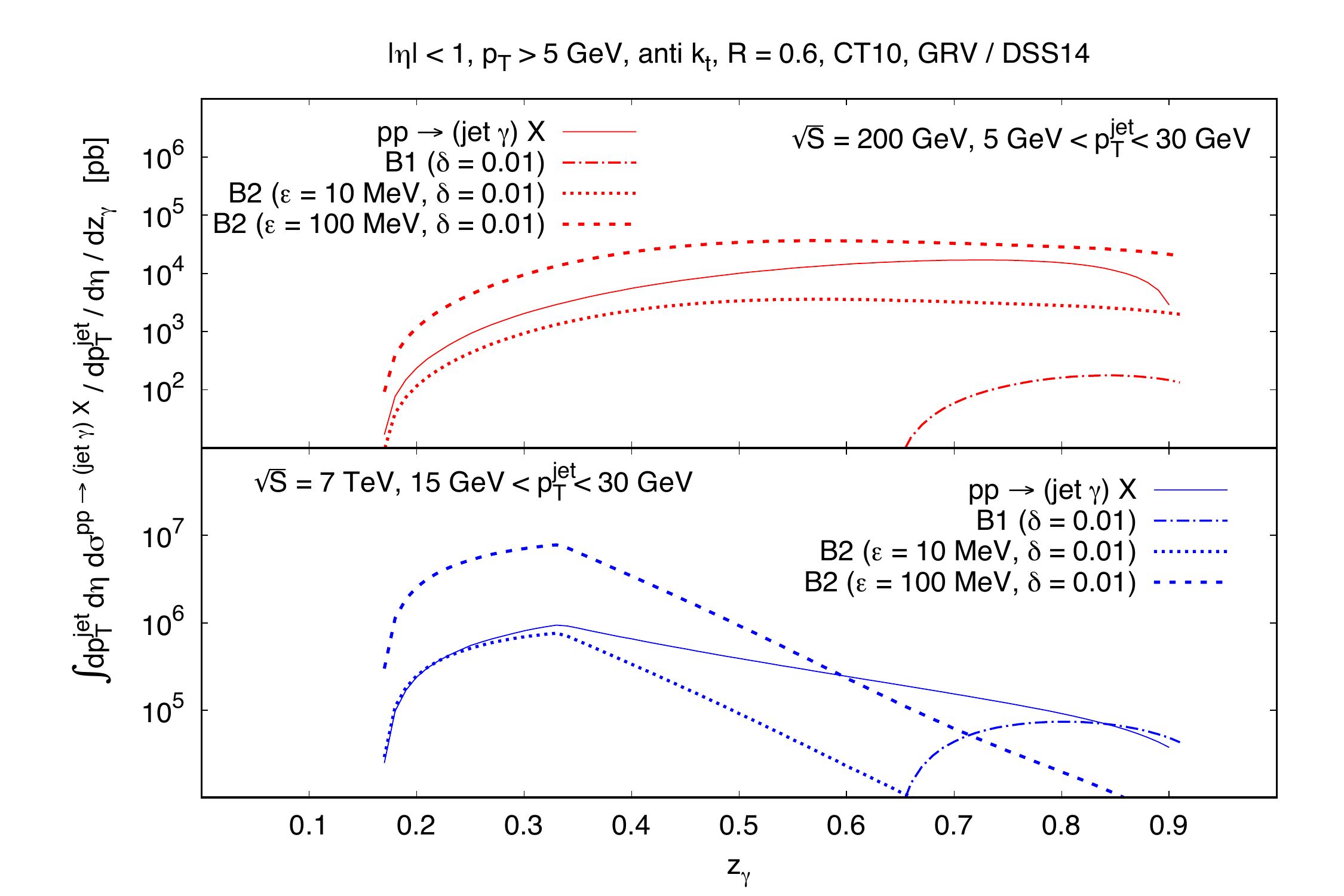}
	\vspace*{-0.5cm}
	\caption{{\sl NLO cross section for $pp\rightarrow (\jet \,\gamma)X$ at $\sqrt{S} = 200 $ GeV 
	as function of $z_\gamma$, for 5 GeV $< p_T^\jet <$ 30 GeV (upper, solid), and at $\sqrt{S} = 7 $ TeV 
	for 15 GeV $< p_T^\jet <$ 30 GeV (lower, solid). The other lines show in both cases the backgrounds
	B1 and B2 from $\pi^0$ decay, for various choices of the detection parameters.}}
	\label{RHIC5-30}
\end{figure}

It is most interesting to consider the cross section differential in $z_\gamma$ as it directly
probes the $z$ dependence of the $D_c^\gamma$. The upper part of Fig.~\ref{RHIC5-30} (solid line) 
shows the cross section for $pp\to(\jet\,\gamma)\,X$ at $\sqrt{S} = 200$ GeV, corresponding to RHIC. 
In the lower part we show results for $pp$ collisions at the LHC at $\sqrt{S} = 7$ TeV.  
We use the CT10 \cite{ct10}  parton distributions and the ``Gl\"uck-Reya-Vogt" (GRV) photon fragmentation 
functions of Ref. \cite{Gluck:1992zx}. The results have been integrated over $|\eta^\jet| < 1$ and over
5 GeV $< p_T^\jet <$ 30 GeV for RHIC and 15 GeV $< p_T^\jet <$ 30 GeV for the LHC. 
We also impose a cut $p_T>5$~GeV on the photon transverse momentum, in order
to make sure that we are in the perturbative regime. As it turns out, this cut also helps to suppress the 
pion decay background (for the LHC the cut causes the ``edges'' in the figure at $z_\gamma=1/3$ 
where 5 GeV$/z_\gamma$ coincides with the lower 
limit of the $p_T^\jet$ integration). The jet is always defined by the anti-$k_t$ algorithm with $R=0.6$. 
We choose $\mu_R = \mu_F = p_T^\jet$  for the renormalization and the initial state factorization scales. 
The final state factorization scale, on the other hand, is set to $\mu_F^{\prime\prime}=p_T^\jet\sqrt{1-z_\gamma}$
(see also~\cite{Chien:2015ctp,Procura:2011aq}).
Along with our value of $R$ this choice helps to minimize the logarithm in Eq.~(\ref{disg}) and hence the direct 
contribution to the cross section. 

In addition to the cross sections we also show the estimated $\pi^0$ background in Fig.~\ref{RHIC5-30}, 
separately for B1 and B2. Here we use the most recent set of pion fragmentation functions of~\cite{dss14}.
We choose the detector resolutions $\delta \phi=\delta \eta\equiv\delta=0.01$ in~(\ref{BG}), which are 
typical values for the RHIC and LHC collider experiments. We furthermore assume two different
values for the energy threshold for photon detection, $\ep=100$~MeV and $\ep=10$~MeV.
We observe that for our choice of kinematics background B1 is small for $\delta \phi=\delta \eta=0.01$, 
since the two photons start to be seen as a single one only when their combined transverse momentum 
exceeds $\sim 20$~GeV. One furthermore sees that background B2 remains below the photon cross section 
only for the relatively low threshold $\ep=10$~MeV. This evidently poses a challenge to experimental studies
of photon-in-jet production.

\begin{figure}[t!]
	\centering
	\includegraphics[width=0.9\textwidth]{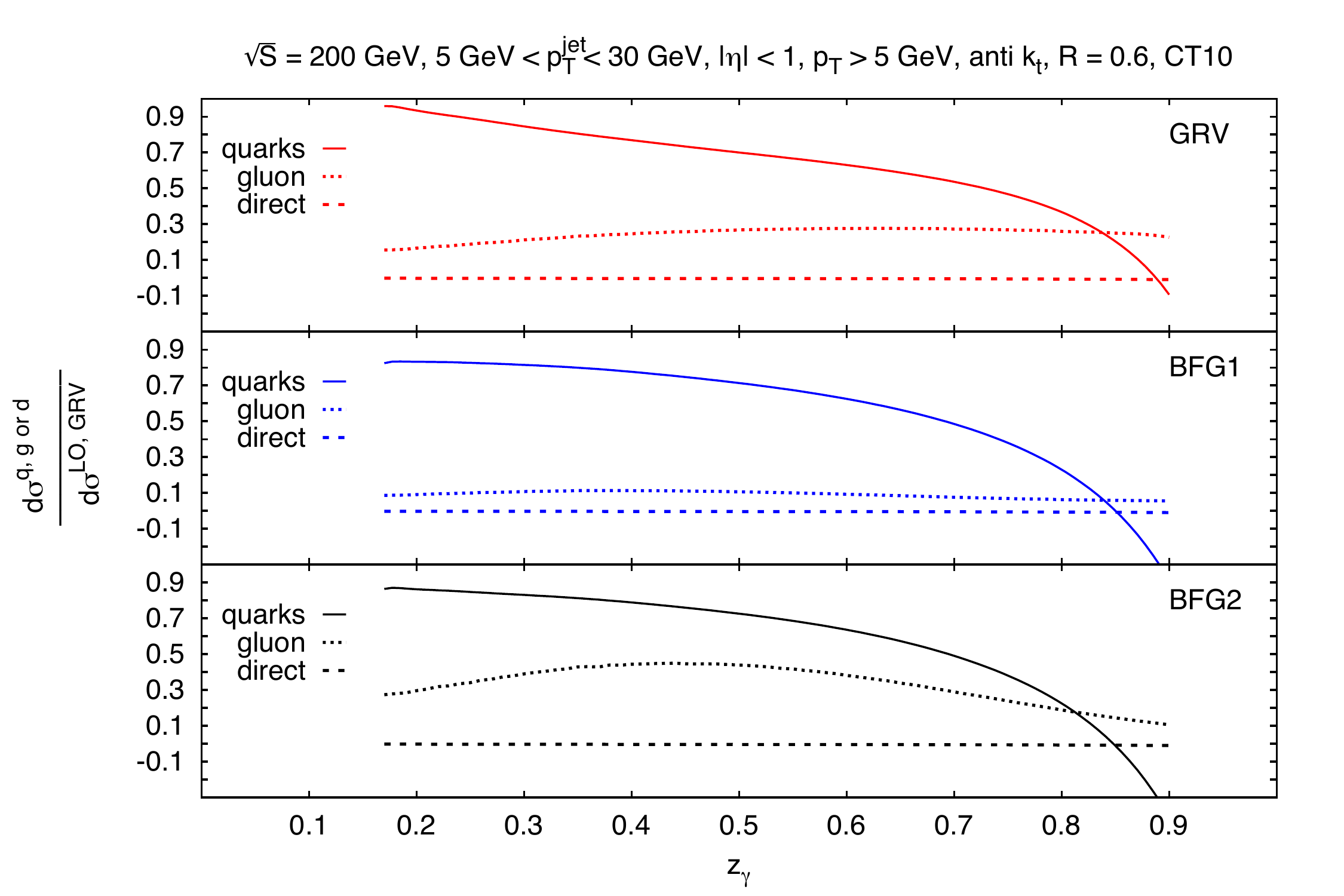}
	\vspace*{-0.5cm}
	\caption{{\sl Normalized quark (solid) and gluon (dotted) fragmentation contributions together with the 
	direct contribution (dashed)  to the 
	cross sections as functions of $z_\gamma$ for 5 GeV $< p_T^\jet <$ 30 GeV at $\sqrt{S} = 200 $ GeV. We 
	show results for the sets of fragmentation functions of Refs.~\cite{Gluck:1992zx,Bourhis:1997yu}.
	For purpose of clearer presentation, all results have been normalized to the LO cross section obtained for 
	GRV fragmentation functions.}}
	\label{qandg}
\end{figure}

Figure~\ref{qandg} analyzes the relative sizes of the contributions of quark/antiquark (summed over all flavors) 
and gluon fragmentation to the cross section, using the RHIC kinematical setup of Fig. \ref{RHIC5-30}. 
We have normalized all contributions to the LO cross section with GRV fragmentation functions. 
We show NLO results for three different sets of photon fragmentation functions: the GRV~\cite{Gluck:1992zx} 
one used so far, and the two sets of ``Bourhis, Fontannaz, Guillet'', Ref.~\cite{Bourhis:1997yu}, which we refer 
to as ``BFG1'' and ``BFG2'' and properly transform to the DIS$_\gamma$ scheme.
One can see that overall gluon fragmentation makes a sizable contribution to the 
photon cross section. The BFG2 set is characterized by an especially large gluon-to-photon fragmentation function, 
which becomes clearly visible in the figure. We also show the direct contribution
based on Eq.~(\ref{disg}) which, in the DIS$_\gamma$ scheme and for our choice of $R$ and $\mu_F^{\prime\prime}$,
is tiny. The cross section for photons produced in jets 
thus offers a direct probe of photon fragmentation also at NLO. 

%%%%%%%%%%%%%%%%%%%%%%%%%%%%%%%%%%%%%%%%%%%%%%%%%%%%%%%
%%%%%%%%%%%%%%%%%%%%%%%%%%%%%%%%%%%%%%%%%%%%%%%%%%%%%%%
\section{Conclusions \label{Concl}}
We have proposed a method of accessing the elusive photon fragmentation functions
in hadronic collisions at RHIC or the LHC. The idea is to identify photons produced inside fully reconstructed jets, 
with the photon treated as part of the jet. The variable $z_\gamma=p_T/p_T^\jet$ introduced in Eq.~(\ref{defzgamma}) 
allows to map out the $z$-dependence of the photon fragmentation functions. We have presented numerical
results for the corresponding cross section that demonstrate the sensitivity to the fragmentation functions.
We have also performed detailed estimates of the background from neutral-pion decay. These suggest that the
measurement should be feasible provided a low photon detection threshold can be chosen. This will 
evidently present a significant challenge; on the other hand, there are likely further techniques
available in experiment that allow to suppress the background, such as subtraction of the background
using a measured pion-in-jet cross section. We are thus optimistic that clean measurements of the $D_c^\gamma$
should become possible in $pp$ collisions, enabling better theoretical control of collider observables involving 
high energy photons. We finally note that detailed future studies of the interplay of QCD radiation inside
jets with photon production may also lead to new techniques for photon isolation.

\section*{Acknowledgments} 
We acknowledge most useful discussions with C. Gagliardi and Y. Akiba. This work was supported in part by the 
Bundesministerium f\"{u}r Bildung und Forschung (BMBF) grant 05P15VTCA1.

%%%%%%%%%%%%%%%%%%%%%%%%%%%%%%%%%%%%%%%%%%%%%%%%%%%%%%%
%%%%%%%%%%%%%%%%%%%%%%%%%%%%%%%%%%%%%%%%%%%%%%%%%%%%%%%
%%%%%%%%%%%%%%%%%%%%%%%%%%%%%%%%%%%%%%%%%%%%%%%%%%%%%%%
\appendix
\renewcommand{\theequation}{A.\arabic{equation}}
\setcounter{equation}{0} 
\section{Jet Functions}\label{appendix:jetfuncs1}
For convenience we collect all NLO jet functions in the $\overline{{\mathrm{MS}}}$ scheme:\\ 
{\it Jet functions $j$:}
\beeq\label{j}
j_{\gamma\to\gamma}(z,\lambda) &=& \delta(1-z) + \mathcal{O}(\alpha)\,,\nn\\[2mm]
j_{\gamma\to q}(z,\lambda) &=& -\frac{\alpha e_q^2}{2\pi}\left[
P_{q\gamma}(z)\log\left( \lambda^2(1-z)^2\right)+ z(1-z)\right]\,,\nn\\[2mm]
j_{q\to\gamma}(z,\lambda) &=&-\frac{\alpha e_q^2}{2\pi}
\left[P_{\gamma q}(z)\log\left( \lambda^2(1-z)^2\right)+ 
z\right]\,,\nn\\[2mm]
j_{\gamma\to g}(z,\lambda) &=& \mathcal{O}(\alpha\alpha_s)\,,\nn\\[2mm]
j_{g\to\gamma}(z,\lambda) &=& \mathcal{O}(\alpha\alpha_s)\,,\nn\\[2mm]
j_{(c\neq\gamma)\to(e\neq\gamma)}(z,\lambda) &=& \delta(1-z) \delta_{ec} + \frac{\alpha_s}{2\pi} f_{c\to e}(z,\lambda) + \mathcal{O}(\alpha_s^2) + \mathcal{O}(\alpha\alpha_s)\,.
\eeeq
{\it Jet functions $\tilde{j}$:}
\beeq\label{jtilde}
\tilde{j}_{\gamma\to\gamma}(z,\lambda) &=& \delta(1-z) + \mathcal{O}(\alpha)\,,\nn\\[2mm]
\tilde{j}_{\gamma\to q}(z,\lambda) &=& \frac{\alpha e_q^2}{2\pi}\left[
P_{q\gamma}(z)\log\left(\lambda^2(1-z)^2\right)+ z(1-z) + \mathcal{I}_{q\gamma}^\text{algo}(z)
\right]\,,\nn\\[2mm]
\tilde{j}_{q\to\gamma}(z,\lambda) &=& \frac{\alpha e_q^2}{2\pi}\left[P_{\gamma q}(z)\log\left( \lambda^2(1-z)^2\right)+ z 
+ \mathcal{I}_{\gamma q}^\text{algo}(z)\right]\,,\nn\\[2mm]
\tilde{j}_{\gamma\to g}(z,\lambda) &=& \mathcal{O}(\alpha\alpha_s)\,,\nn\\[2mm]
\tilde{j}_{g\to\gamma}(z,\lambda) &=& \mathcal{O}(\alpha\alpha_s)\,,\nn\\[2mm]
\tilde{j}_{(e\neq\gamma)\to(c^\prime \neq\gamma)}(z,\lambda) &=& \delta(1-z) \delta_{ec^\prime} + \frac{\alpha_s}{2\pi} \tilde{f}_{e\to c^\prime}(z,\lambda) + \mathcal{O}(\alpha_s^2) + \mathcal{O}(\alpha\alpha_s)\,,
\eeeq
with
\beq\label{Ialg}
\mathcal{I}_{c^\prime c}^\text{algo}(z) = \left\{ \begin{array}{cl} 2 P_{c'c}(z) 
\log\left(\frac{z}{1-z}\right)\Theta(1/2-z) & {\mathrm{cone\;algorithm}}\,,\\[2mm]
2 P_{c'c}(z) \log z & {\mathrm{(anti-)}}k_t {\mathrm{\;algorithm}}\,,\\[2mm]
P_{c'c}(z) \left[\log(z) + \log\left(\frac{z}{1-z}\right)\Theta(1/2-z)\right] & J_{E_T} {\mathrm{\;algorithm}}\,.
\end{array} \right.
\eeq
The last lines in each of Eqs.~\eqref{j} and \eqref{jtilde} describe the ``pure-QCD'' type jet functions without any 
photonic contributions. They are shown only for the sake of completeness and listed explicitly in \cite{Kaufmann:2015hma},
from which the functions $f_{c\to e}(z,\lambda)$ and $\tilde{f}_{e\to c^\prime}(z,\lambda)$ may be read off easily.

The expressions for $\tilde{j}_{q\to\gamma}(z,\lambda)$ in the DIS$_\gamma$ scheme may be 
obtained via the scheme transformation as shown in Eq.~(\ref{disg}). All other jet functions are
unaffected by that transformation.


\newpage
\begin{thebibliography}{99}

%\cite{CMS:2015dxe}
\bibitem{diphoton}
%\bibitem{CMS:2015dxe} 
  CMS Collaboration,
  %``Search for new physics in high mass diphoton events in proton-proton
%  collisions at 13TeV,''
  CMS-PAS-EXO-15-004;
  %%CITATION = CMS-PAS-EXO-15-004;%%
  %294 citations counted in INSPIRE as of 22 Apr 2016
  %\cite{}
%\bibitem{} 
ATLAS collaboration,
  %``Search for resonances decaying to photon pairs in 3.2 fb$^{-1}$ of $pp$ collisions at $\sqrt{s}$ = 13 TeV with the ATLAS detector,''
  ATLAS-CONF-2015-081.
  %%CITATION = ATLAS-CONF-2015-081;%%
  %304 citations counted in INSPIRE as of 22 Apr 2016

\bibitem{Owens:1986mp} 
  J.~F.~Owens,
  %``Large Momentum Transfer Production of Direct Photons, Jets, and Particles,''
  Rev.\ Mod.\ Phys.\  {\bf 59}, 465 (1987).
%  doi:10.1103/RevModPhys.59.465
  %%CITATION = doi:10.1103/RevModPhys.59.465;%%
  %540 citations counted in INSPIRE as of 25 Dec 2015

\bibitem{singlephoton} 
  P.~Aurenche, A.~Douiri, R.~Baier, M.~Fontannaz and D.~Schiff,
  %``Prompt Photon Production at Large p(T) in QCD Beyond the Leading Order,''
  Phys.\ Lett.\ B {\bf 140}, 87 (1984);
  P.~Aurenche, R.~Baier, M.~Fontannaz and D.~Schiff,
  %``Prompt Photon Production at Large p(T) Scheme Invariant QCD Predictions and Comparison with Experiment,''
  Nucl.\ Phys.\ B {\bf 297}, 661 (1988);
 H.~Baer, J.~Ohnemus and J.~F.~Owens,
  %``A Next-to-leading Logarithm Calculation of Direct Photon Production,''
  Phys.\ Rev.\ D {\bf 42}, 61 (1990); 
  L.~E.~Gordon and W.~Vogelsang,
  %``Polarized and unpolarized prompt photon production beyond the leading order,''
  Phys.\ Rev.\ D {\bf 48}, 3136 (1993);
  L.~E.~Gordon and W.~Vogelsang,
  %``Polarized and unpolarized isolated prompt photon production beyond the leading order,''
  Phys.\ Rev.\ D {\bf 50}, 1901 (1994).

\bibitem{singleparton} 
  F.~Aversa, P.~Chiappetta, M.~Greco and J.~P.~Guillet,
  %``QCD Corrections to Parton-Parton Scattering Processes,''
  Nucl.\ Phys.\ B {\bf 327}, 105 (1989);
D.~de Florian,
  %``Next-to-leading order QCD corrections to one hadron production in polarized pp collisions at RHIC,''
  Phys.\ Rev.\ D {\bf 67}, 054004 (2003)  [hep-ph/0210442];
  B.~J\"{a}ger, A.~Sch\"{a}fer, M.~Stratmann and W.~Vogelsang,
  %``Next-to-leading order QCD corrections to high p(T) pion production in longitudinally polarized pp collisions,''
  Phys.\ Rev.\ D {\bf 67}, 054005 (2003)  [hep-ph/0211007].

\bibitem{resu} 
E.~Laenen, G.~Oderda and G.~F.~Sterman,
  %``Resummation of threshold corrections for single particle inclusive cross-sections,''
  Phys.\ Lett.\ B {\bf 438}, 173 (1998)
%  doi:10.1016/S0370-2693(98)00960-5
  [hep-ph/9806467];
S.~Catani, M.~L.~Mangano and P.~Nason,
  %``Sudakov resummation for prompt photon production in hadron collisions,''
  JHEP {\bf 9807}, 024 (1998)
%  doi:10.1088/1126-6708/1998/07/024
  [hep-ph/9806484]; 
  S.~Catani, M.~L.~Mangano, P.~Nason, C.~Oleari and W.~Vogelsang,
  %``Sudakov resummation effects in prompt photon hadroproduction,''
  JHEP {\bf 9903}, 025 (1999)
%  doi:10.1088/1126-6708/1999/03/025
  [hep-ph/9903436];  
N.~Kidonakis and J.~F.~Owens,
  %``Soft gluon resummation and NNLO corrections for direct photon production,''
  Phys.\ Rev.\ D {\bf 61}, 094004 (2000)
%  doi:10.1103/PhysRevD.61.094004
  [hep-ph/9912388];  
 G.~F.~Sterman and W.~Vogelsang,
  %``Threshold resummation and rapidity dependence,''
  JHEP {\bf 0102}, 016 (2001)
%  doi:10.1088/1126-6708/2001/02/016
  [hep-ph/0011289];  
 E.~Laenen, G.~F.~Sterman and W.~Vogelsang,
  %``Higher order QCD corrections in prompt photon production,''
  Phys.\ Rev.\ Lett.\  {\bf 84}, 4296 (2000)
%  doi:10.1103/PhysRevLett.84.4296 
  [hep-ph/0002078];  
 P.~Bolzoni, S.~Forte and G.~Ridolfi,
  %``Renormalization group approach to Sudakov resummation in prompt photon production,''
  Nucl.\ Phys.\ B {\bf 731}, 85 (2005)
%  doi:10.1016/j.nuclphysb.2005.07.036
  [hep-ph/0504115]; 
  D.~de Florian and W.~Vogelsang,
  %``Threshold resummation for the prompt-photon cross section revisited,''
  Phys.\ Rev.\ D {\bf 72}, 014014 (2005)
%  doi:10.1103/PhysRevD.72.014014
  [hep-ph/0506150];
  T.~Becher and M.~D.~Schwartz,
  %``Direct photon production with effective field theory,''
  JHEP {\bf 1002}, 040 (2010)
%  doi:10.1007/JHEP02(2010)040
  [arXiv:0911.0681 [hep-ph]].
  
\bibitem{Buskulic:1995au} 
  D.~Buskulic {\it et al.} [ALEPH Collaboration],
  %``First measurement of the quark to photon fragmentation function,''
  Z.\ Phys.\ C {\bf 69}, 365 (1996);
%  doi:10.1007/BF02907417  
  K.~Ackerstaff {\it et al.} [OPAL Collaboration],
  %``Measurement of the quark to photon fragmentation function through the inclusive production of prompt photons in hadronic Z0 decays,''
  Eur.\ Phys.\ J.\ C {\bf 2}, 39 (1998)
%  doi:10.1007/s100520050122
  [hep-ex/9708020].

\bibitem{Aurenche:1992yc} 
  P.~Aurenche, P.~Chiappetta, M.~Fontannaz, J.~P.~Guillet and E.~Pilon,
  %``Next-to-leading order bremsstrahlung contribution to prompt photon production,''
  Nucl.\ Phys.\ B {\bf 399}, 34 (1993).
  
\bibitem{Gluck:1992zx} 
  M.~Gl\"{u}ck, E.~Reya and A.~Vogt,
  %``Parton fragmentation into photons beyond the leading order,''
  Phys.\ Rev.\ D {\bf 48}, 116 (1993)
  [Phys.\ Rev.\ D {\bf 51}, 1427 (1995)].

\bibitem{Bourhis:1997yu} 
  L.~Bourhis, M.~Fontannaz and J.~P.~Guillet,
  %``Quarks and gluon fragmentation functions into photons,''
  Eur.\ Phys.\ J.\ C {\bf 2}, 529 (1998)
  [hep-ph/9704447].
  %%CITATION = HEP-PH/9704447;%%

\bibitem{GehrmannDeRidder:1997wx} 
  A.~Gehrmann-De Ridder, T.~Gehrmann and E.~W.~N.~Glover,
  %``Radiative corrections to the photon + 1 jet rate at LEP,''
  Phys.\ Lett.\ B {\bf 414}, 354 (1997)
%  doi:10.1016/S0370-2693(97)01186-6
  [hep-ph/9705305];
A.~Gehrmann-De Ridder and E.~W.~N.~Glover,
  %``A Complete O (alpha alpha-s) calculation of the photon + 1 jet rate in e+ e- annihilation,''
  Nucl.\ Phys.\ B {\bf 517}, 269 (1998)
%  doi:10.1016/S0550-3213(97)00818-3
  [hep-ph/9707224].  

\bibitem{Vogelsang:1995bg} 
  W.~Vogelsang and A.~Vogt,
  %``Constraints on the proton ' $s$ gluon distribution from prompt photon production,''
  Nucl.\ Phys.\ B {\bf 453}, 334 (1995)
%  doi:10.1016/0550-3213(95)00424-Q
  [hep-ph/9505404].

\bibitem{Gluck:1994iz} 
  M.~Gl\"{u}ck, L.~E.~Gordon, E.~Reya and W.~Vogelsang,
  %``High P($T$) photon production at $p \bar{p}$ collider,''
  Phys.\ Rev.\ Lett.\  {\bf 73}, 388 (1994).
  %  doi:10.1103/PhysRevLett.73.388
  %%CITATION = doi:10.1103/PhysRevLett.73.388;%%    
  
       %\cite{Klasen:2014xfa}
             \bibitem{Klasen:2014xfa} 
                  M.~Klasen and F.~König,
                  %``New information on photon fragmentation functions,?'
                  Eur.\ Phys.\ J.\ C {\bf 74}, no. 8, 3009 (2014)
                 % doi:10.1140/epjc/s10052-014-3009-x
                 [arXiv:1403.2290 [hep-ph]].
                 %%CITATION = doi:10.1140/epjc/s10052-014-3009-x;%%
                 %3 citations counted in INSPIRE as of 10 Jun 2016
  
\bibitem{Kaufmann:2015hma} 
T.~Kaufmann, A.~Mukherjee and W.~Vogelsang,
  %``Hadron Fragmentation Inside Jets in Hadronic Collisions,''
  Phys.\ Rev.\ D {\bf 92}, 054015 (2015)
%  doi:10.1103/PhysRevD.92.054015
  [arXiv:1506.01415 [hep-ph]].
  
\bibitem{Arleo:2013tya} 
  F.~Arleo, M.~Fontannaz, J.~P.~Guillet and C.~L.~Nguyen,
  %``Probing fragmentation functions from same-side hadron-jet momentum correlations in p-p collisions,''
  JHEP {\bf 1404}, 147 (2014)
%  doi:10.1007/JHEP04(2014)147
  [arXiv:1311.7356 [hep-ph]].
  
\bibitem{Chien:2015ctp} 
  Y.~T.~Chien, Z.~B.~Kang, F.~Ringer, I.~Vitev and H.~Xing,
  %``Jet fragmentation functions in proton-proton collisions using soft-collinear effective theory,''
  arXiv:1512.06851 [hep-ph].    
  
 %\cite{Procura:2011aq}
\bibitem{Procura:2011aq} 
  M.~Procura and W.~J.~Waalewijn,
  %``Fragmentation in Jets: Cone and Threshold Effects,''
  Phys.\ Rev.\ D {\bf 85}, 114041 (2012)
%  doi:10.1103/PhysRevD.85.114041
  [arXiv:1111.6605 [hep-ph]].
  %%CITATION = doi:10.1103/PhysRevD.85.114041;%%
  %18 citations counted in INSPIRE as of 22 Apr 2016
  
\bibitem{Belghobsi:2009hx} 
  Z.~Belghobsi, M.~Fontannaz, J.-P.~Guillet, G.~Heinrich, E.~Pilon and M.~Werlen,
  %``Photon - Jet Correlations and Constraints on Fragmentation Functions,''
  Phys.\ Rev.\ D {\bf 79}, 114024 (2009)
%  doi:10.1103/PhysRevD.79.114024
  [arXiv:0903.4834 [hep-ph]].
    
\bibitem{Glover:1993xc} 
  E.~W.~N.~Glover and A.~G.~Morgan,
  %``Measuring the photon fragmentation function at LEP,''
  Z.\ Phys.\ C {\bf 62}, 311 (1994).
      
\bibitem{GehrmannDeRidder:2006vn} 
A.~Gehrmann-De Ridder, G.~Kramer and H.~Spiesberger,
  %``Photon fragmentation in large Q**2 e p collisions at next-to-leading order QCD,''
  Eur.\ Phys.\ J.\ C {\bf 11}, 137 (1999)
%  doi:10.1007/s100520050620
  [hep-ph/9907511]; Nucl.\ Phys.\ B {\bf 578}, 326 (2000)
%  doi:10.1016/S0550-3213(00)00228-5
  [hep-ph/0003082];
  A.~Gehrmann-De Ridder, T.~Gehrmann and E.~Poulsen,
  %``Measuring the Photon Fragmentation Function at HERA,''
  Eur.\ Phys.\ J.\ C {\bf 47}, 395 (2006)
% doi:10.1140/epjc/s2006-02574-x
  [hep-ph/0604030].
  
 \bibitem{algo:cone} 
  G.~P.~Salam and G.~Soyez,
  %``A Practical Seedless Infrared-Safe Cone jet algorithm,''
  JHEP {\bf 0705}, 086 (2007)
  [arXiv:0704.0292 [hep-ph]].
  
\bibitem{algo:kt} 
  S.~D.~Ellis and D.~E.~Soper,
  %``Successive combination jet algorithm for hadron collisions,''
  Phys.\ Rev.\ D {\bf 48}, 3160 (1993)
  [hep-ph/9305266];
  S.~Catani, Y.~L.~Dokshitzer, M.~H.~Seymour and B.~R.~Webber,
  %``Longitudinally invariant $K_t$ clustering algorithms for hadron hadron collisions,''
  Nucl.\ Phys.\ B {\bf 406}, 187 (1993);
  M.~Cacciari, G.~P.~Salam and G.~Soyez,
  %``The Anti-k(t) jet clustering algorithm,''
  JHEP {\bf 0804}, 063 (2008)
  [arXiv:0802.1189 [hep-ph]].
  
\bibitem{algo:JET} 
  H.~Georgi,
  %``A Simple Alternative to Jet-Clustering Algorithms,''
  arXiv:1408.1161 [hep-ph];
  Y.~Bai, Z.~Han and R.~Lu,
  %``$J_{E_T}$: a global jet finding algorithm,''
  JHEP {\bf 1503}, 102 (2015)
  [arXiv:1411.3705 [hep-ph]];
  J.~Thaler,
  %``Jet maximization, axis minimization, and stable cone finding,''
  Phys.\ Rev.\ D {\bf 92}, no. 7, 074001 (2015)
%  doi:10.1103/PhysRevD.92.074001
  [arXiv:1506.07876 [hep-ph]].
  
\bibitem{Jager:2004jh} 
  B.~J\"{a}ger, M.~Stratmann and W.~Vogelsang,
  %``Single inclusive jet production in polarized $p p$ collisions at $O(alpha^3_s)$,''
  Phys.\ Rev.\ D {\bf 70}, 034010 (2004)
  [hep-ph/0404057]; 
%\bibitem{Mukherjee:2012uz} 
  A.~Mukherjee and W.~Vogelsang,
  %``Jet production in (un)polarized pp collisions: dependence on jet algorithm,''
  Phys.\ Rev.\ D {\bf 86}, 094009 (2012)
  [arXiv:1209.1785 [hep-ph]];
%\bibitem{Kaufmann:2014nda} 
  T.~Kaufmann, A.~Mukherjee and W.~Vogelsang,
  %``Next-to-leading Order Calculation for Jets Defined by a Maximized Jet Function,''
  Phys.\ Rev.\ D {\bf 91}, 034001 (2015)
  [arXiv:1412.0298 [hep-ph]].

\bibitem{Dasgupta:2016bnd} 
  M.~Dasgupta, F.~A.~Dreyer, G.~P.~Salam and G.~Soyez,
  %``Inclusive jet spectrum for small-radius jets,''
  arXiv:1602.01110 [hep-ph].
    
\bibitem{Gluck:1991ee} 
  M.~Gl\"{u}ck, E.~Reya and A.~Vogt,
  %``Parton structure of the photon beyond the leading order,''
  Phys.\ Rev.\ D {\bf 45}, 3986 (1992).
%  doi:10.1103/PhysRevD.45.3986 
   
\bibitem{Agashe:2014kda} 
  K.~A.~Olive {\it et al.} [Particle Data Group Collaboration],
  %``Review of Particle Physics,''
  Chin.\ Phys.\ C {\bf 38}, 090001 (2014).
  
\bibitem{Adare:2012yt} 
  A.~Adare {\it et al.} [PHENIX Collaboration],
  %``Direct-Photon Production in $p+p$ Collisions at $\sqrt{s}=200$ GeV at Midrapidity,''
  Phys.\ Rev.\ D {\bf 86}, 072008 (2012)
%  doi:10.1103/PhysRevD.86.072008
  [arXiv:1205.5533 [hep-ex]]; Y.~Akiba, {\it private communications}.
  
\bibitem{carl} C. Gagliardi, {\it private communications}.

\bibitem{WZW} 
  J.~Wess and B.~Zumino,
  %``Consequences of anomalous Ward identities,''
  Phys.\ Lett.\ B {\bf 37}, 95 (1971).
 % doi:10.1016/0370-2693(71)90582-X;
  
\bibitem{ct10} 
  H.~L.~Lai, M.~Guzzi, J.~Huston, Z.~Li, P.~M.~Nadolsky, J.~Pumplin and C.-P.~Yuan,
  %``New parton distributions for collider physics,''
  Phys.\ Rev.\ D {\bf 82}, 074024 (2010)
  [arXiv:1007.2241 [hep-ph]].
  %%CITATION = ARXIV:1007.2241;%%
  
\bibitem{dss14} 
  D.~de Florian, R.~Sassot, M.~Epele, R.~J.~Hern\'{a}ndez-Pinto and M.~Stratmann,
  %``Parton-to-Pion Fragmentation Reloaded,''
  Phys.\ Rev.\ D {\bf 91}, 014035 (2015)
  [arXiv:1410.6027 [hep-ph]].
  %%CITATION = ARXIV:1410.6027;%%
  
\end{thebibliography}
\end{document}